\documentclass[doublecol]{epl2}

\newcommand{\imi}{{\mbox{i}}}
\newcommand{\Imm}{{\Im \mbox{m}}}

\title{Solving the Coulomb scattering problem using 
the complex scaling method}

\author{M. V. Volkov\inst{1,2}\thanks{E-mail: \email{miha@physto.se}}
N. Elander\inst{1} E. Yarevsky\inst{2} \and S. L. Yakovlev\inst{2}}
\shortauthor{M. V. Volkov\etal}

\institute{
  \inst{1} Department of Physics, AlbaNova University Center,
Stockholm University, 106 91 Stockholm, Sweden \\
  \inst{2} Department of Computational Physics, St Petersburg
State University, 198504 St Petersburg, Russia
}

\abstract{Based on the work of Nuttall and Cohen [Phys. Rev. {\bf 188} (1969) 1542]
and Resigno \etal{} [Phys. Rev. A {\bf 55} (1997) 4253] we present a rigorous
formalism for solving the scattering problem for long-range interactions without
using exact asymptotic boundary conditions. The long-range interaction may contain
both Coulomb and short-range potentials. The exterior complex scaling method,
applied to a specially constructed inhomogeneous Schr\"odinger equation, transforms
the scattering problem into a boundary problem with zero boundary conditions.
The local and integral representations for the scattering amplitudes have been
derived. The formalism is illustrated with numerical examples.}

\begin{document}
\maketitle


\date{\today}

\section{Introduction}
Few-body systems held together by a mutual Coulomb interaction are of great
interest in many areas of quantum physics. However, solving the Coulomb scattering
problem is a very difficult task both from the theoretical as well as the
computational points of view due to the long-range character of the Coulomb
interaction. The asymptotic boundary conditions for the wave function at large
separations between particles are already complicated for the few-body scattering
problem with short-range interactions~\cite{Fadd-Merk}. They become even more
complicated for the long-range case when the Coulomb potential is present in the
interaction~\cite{Merk}. Therefore, a method which allows the problem to be solved
without explicit use of the asymptotic form of the wave function is of great
interest from both the theoretical and computational points of view.

One of such methods was proposed by Nuttal and Cohen \cite{NutCoh}.
The approach is based on the complex scaling theory~\cite{r:BalslevCombes71}.
The idea can briefly be formulated as follows.

The Schr\"odinger equation is recast into its inhomogeneous (driven) form by
splitting the wave function into the sum $\Psi=\Psi_{in}+\Psi_{sc}$ of the
incident $\Psi_{in}$ and scattered $\Psi_{sc}$ waves as 
\begin{equation}
(H_0+V-E)\Psi_{sc}= -V\Psi_{in}.
\label{DSE}
\end{equation}
The scattered wave is the subject of the purely outgoing boundary condition
$\Psi_{sc}\propto \exp\{ \imi \sqrt{E}r\}$. Then, after the complex scaling
transformation, the original real coordinate will be replaced by $z$,
$\Imm \, z>0$, implying that the scattered wave vanishes exponentially
$\Psi_{sc}\propto \exp\{ \imi\, \sqrt{E}z\}$ as $|z|\to \infty$.
This property is consistent with Eq.~(\ref{DSE}) as long as the right hand
side term $V(z)\Psi_{in}(z)$ vanishes as $|z|\to \infty$. This condition is
always fulfilled if the potential has only finite range. Otherwise only
exponentially decreasing potentials $V(r)\propto \exp\{-\mu r \}$ are
admissible~\cite{NutCoh}. The problem is that the incident wave $\Psi_{in}$
always contains the incoming wave $\exp\{-\imi\sqrt{E}r \}$ in addition to the
outgoing wave. This wave increases in magnitude after the complex scaling as
$|z|\to \infty$. This increase can only be compensated for if the potential
decreases exponentially with increasing $z$.

In the two body scattering problem the Coulomb potential can be implemented into
the approach just discussed if it is included in the Hamiltonian $H_0$, while $V$
assumes the short-range part of the interaction. In this case the incident wave
$\Psi_{in}$ is represented by the Coulomb wave function, which is known analytically.
This approach has been successfully used for calculations in atomic~\cite{McCurdy2004}
and nuclear~\cite{Kruppa} physics. Unlike the two body case, the analytic solution
for the Coulomb problem does not exist if three or more particles are involved in
the scattering process. Therefore, this version of the method does not work for
more than two charged particles.

The mathematics is only valid for potentials $V$, which at large distances do not
decrease slower than exponentially. Nevertheless, it was found by McCurdy, Rescigno
and coworkers~\cite{number1,McCurdy2004,e-impact} that the method can give good
results for the slow decreasing potentials when applied in the following way.
Let $R$ be a parametrical radius chosen such that $V(r)\ll E$ for $r\geq R$.
Then Eq.~(\ref{DSE}) can be approximated by 
\begin{equation}
(H_0+V-E)\Psi_{sc}=-V_{R}\Psi_{in}.
\label{DRE_R}
\end{equation}
Here the finite-range potential $V_R$ is introduced in such a way that $V_{R}=V$,
if $r\leq R$, and $V_R=0$ otherwise. The exterior complex scaling
(ECS)~\cite{r:SimonECS79,r:HislopSigal86} is then applied to Eq.~(\ref{DRE_R})
instead of Eq.~(\ref{DSE}). The mathematical properties of this interpretation of
the method have not been studied earlier to the best of our knowledge.
Furthermore, its extension to the Coulomb case remained questionable.
However, the definite success in applying this method to important and complicated
problems (see papers \cite{McCurdy2004,e-impact}, \cite{R:toprev} and references
therein) makes a detailed study urgent.

The aim of the present note is to set the above formulated approximative approach
on a mathematically solid basis. A new formalism for solving the scattering problem
with long-range interactions, including the Coulomb potential, is presented.
The new formalism is based on an inhomogeneous Schr\"odinger equation with the
right hand side containing the finite range potential $V_R$ as in Eq.~(\ref{DRE_R}).
The present contribution will open the way to further rigorous extensions of the
scattering problem with more than two particles.

In this study we consider the scattering problem for a two body system with a
central potential in the center of mass frame.
Therefore, the Schr\"odinger equation for one partial wave $\Psi_{\ell}(k,r)$, with
angular momentum $\ell$ and momentum $k=\sqrt{E}$, is considered.


\section{Scattering problem and driven Schr{\"o}dinger equation\label{rsp}}
Let $\Psi_{\ell}$ be the wave function for a given orbital momentum $\ell$.
It is the solution to the partial wave Schr\"odinger equation 
$ \left( H_\ell+V-k^2\right)\Psi_\ell=0 $
with the ``free" Hamiltonian $H_\ell=-\partial^2_r+\ell(\ell+1)/r^2$ and the
interaction potential $V(r)=2k\eta/r + V_{s}(r)$. $V_s$ is assumed to vanish faster
than $1/r^2$ at large distances, and is referred to as the short-range potential
in this note.

The scattering solution is defined by the boundary condition $\Psi_\ell(k,0)=0$
and the asymptotics as $r\to \infty$ yielding
\begin{equation}
\Psi_\ell(k,r) \sim  e^{\imi \sigma_{\ell}}F_{\ell}(\eta,kr)+ {\cal A}\, u^{+}_{\ell}(\eta,kr).
\label{Psi-as}
\end{equation}
Here $u^{\pm}_\ell=e^{\mp \imi\sigma_\ell}(G_\ell\pm \imi F_\ell)$, where
$F_\ell(G_\ell)$ is the regular (irregular) Coulomb wave function~\cite{Abramowitz},
and $\sigma_\ell=\arg\Gamma(1+\ell+\imi\eta)$ is the Coulomb phase shift.
The scattering amplitude
\begin{equation}
{\cal A}=e^{2\imi\sigma_\ell}\frac{e^{2\imi \delta_\ell}-1}{2\imi}
\label{f}
\end{equation}
is determined by the phase shift $\delta_\ell$, which is due to the presence of
the potential $V_s$.

In order to reformulate the problem in terms of a driven Schr\"odinger equation,
the potential $V$ is split into the sum $V=V_R+V^R$ containing the just introduced
interior $V_R$ part and the exterior $V^R$ part. The splitting of the wave function
$\Psi_\ell=\Psi_R+\Psi^R$ leads to the driven equation 
\begin{equation}
(H_\ell+V-k^2)\Psi_R=-V_R\Psi^R
\label{DrivenSE}
\end{equation}
provided that $\Psi^R$ obeys the Schr\"odinger equation with the exterior potential
\begin{equation}
(H_\ell+V^{R}-k^2)\Psi^R=0.
\label{SE-R}
\end{equation}
The solution to Eq.~(\ref{SE-R}) is constructed in such a way that it incorporates
the incident wave $e^{\imi\sigma_\ell}F_\ell$. Therefore, $\Psi^R$ has to represent
the regular solution at $r=0$ such that $\Psi^R(k,0)=0$ and has to fulfill the
asymptotics that is similar to (\ref{Psi-as})
\begin{equation}
\Psi^R(k,r) \sim e^{\imi \sigma_\ell}F_{\ell}(\eta,kr)+ {\cal A}^R\, u^{+}_{\ell}(\eta,kr)
\label{Psi^R-as}
\end{equation}
with the amplitude ${\cal A}^R$ given by
\begin{equation}
{\cal A}^R=e^{2\imi\sigma_\ell}\frac{e^{2\imi \delta^R}-1}{2\imi}.
\label{f^R}
\end{equation}
Note that the asymptotics (\ref{Psi^R-as}) is relevant to that part of the function
$\Psi^R(k,r)$, which is defined for $r>R$. For $r\leq R$ the potential $V^R$
vanishes ($V^R(r)=0$). Therefore, the function $\Psi^{R}(k,r)$ must be proportional
to the Riccati-Bessel function ${\hat {j_\ell}}$~\cite{Abramowitz}:
\begin{equation}
\Psi^{R}(k,r)=a^R {\hat {j_\ell}}(kr).
\label{Psi^R<R}
\end{equation}
In order to determine the function $\Psi^R(k,r)$ for $r>R$ it is useful to rewrite
the asymptotics (\ref{Psi^R-as}) in terms of Coulomb functions $u^{\pm}_{\ell}$
\begin{equation}
\Psi^R(k,r)\sim \frac{1}{2\imi}\left[ -u^{-}_{\ell}(\eta,kr) +
   S^R u^{+}_{\ell}(\eta,kr)\right],
\label{u-asymp}
\end{equation}
where $S^R=e^{2\imi(\sigma_{\ell}+\delta^R)}$. Then the function $\Psi^R(k,r)$ is expressed as
\begin{equation}
\Psi^R(k,r)=\frac{1}{2\imi}\left[ -{\hat U}^R_{-}(k,r)+S^R {\hat U}^R_{+}(k,r) \right]
\label{PsiUU}
\end{equation}
in terms of Jost solutions ${\hat U}^R_{\pm}$. The latter are the solutions to the Volterra integral equations~\cite{Newton}
\begin{equation}
{\hat U}^R_{\pm}(k,r)=u^{\pm}_{\ell}(\eta,kr)-\int_{r}^{\infty} dr'\, G^{C}_{\ell}(r,r',k)V_{s}(r'){\hat U}^R_{\pm}(k,r').
\label{UJost}
\end{equation}
The kernel $G^{C}_{\ell}(r,r',k)$  here 
is given by
\begin{equation}
\begin{array}{l}
G^{C}_{\ell}(r,r',k)=  \\
\frac{\imi}{2k}\left[ u^+_{\ell}(\eta,kr)u^-_{\ell}(\eta,kr')-u^+_{\ell}(\eta,kr')u^-_{\ell}(\eta,kr)
\right].
\end{array}
\label{GC}
\end{equation}
The requirement for the wave function and its derivative to obey the continuity
conditions $\partial^m \Psi^R(k,R-0)=\partial^m\Psi^{R}(k,R+0)$  $(m=0,1)$  at
the point $r=R$ completes the construction of $\Psi^R$. This construction provides
a way to calculate $a^R$ and ${S}^R$
\begin{equation}
\begin{array}{cl}
-2\imi a^R& =
W_R({\hat U}^R_{-},\,{\hat U}^R_{+})/W_{R}({\hat{j_\ell}},\,{\hat U}^R_{+}) , \\
S^R& =W_R({\hat U}^R_{-},\,{\hat{j_\ell}})/W_{R}({{\hat U}^R_{+},\,\hat{j_\ell}}).
\end{array}
\label{aS^R}
\end{equation}
Here $W_R(f,g)$ denotes the Wronskian $f(r)g'(r)-g(r)f'(r)$ calculated at $r=R$.
The expressions for $a^R$ and $S^R$ become simpler when the radius $R$ is chosen
large enough. If the asymptotics ${\hat U}^R_{\pm}(k,R)\sim u^{\pm}_{\ell}(\eta,kR)$
can be used and at the same time $kR\gg \ell(\ell+1)+\eta^2$, then
\begin{equation}
a^R \sim e^{\imi\eta\log2kR},\ \ S^R\sim e^{2\imi \eta\log 2kR}.
\label{aSRas}
\end{equation}
From this representation of $S^R$ we get the phase shift 
asymptotics $\delta^{R}\sim \eta\log 2kR-\sigma_{\ell}$.

Once the wave function $\Psi^R$ has been constructed, Eq.~(\ref{DrivenSE}) is
well defined. By imposing the boundary conditions
\begin{equation}
\begin{array}{l}
\Psi_R(k,0)=0, \\
\Psi_R(k,r) \sim {\cal A}_R \,u^{+}_{\ell}(\eta,kr), \ \ r\to \infty\
\end{array}
\label{BCPsi_R}
\end{equation}
this equation determines the remainder  of the scattering wave function
$\Psi_R=\Psi_{\ell}-\Psi^R$. The amplitude ${\cal A}_R$ is given by
${\cal A}_R={\cal A}-{\cal A}^R$. The representation of the
amplitude ${\cal A}_R$ in terms of the residual phase shift
$\delta_R=\delta_{\ell}-\delta^R$ has the standard form
\begin{equation}
{\cal A}_R=e^{2\imi(\sigma_{\ell}+\delta^R)}\frac{e^{2\imi \delta_R}-1}{2\imi}.
\label{A_R}
\end{equation}

The structure of the formula (\ref{Psi^R<R}) suggests a further simplification.
If a new function $\Phi_R$ is introduced by $\Phi_{R}=(a^{R})^{-1}\Psi_R$, then
Eq.~(\ref{DrivenSE})
transforms into
\begin{equation}
(H_\ell+V(r)-k^2)\Phi_{R}(k,r)=-V_R(r){\hat{j_\ell}(kr)}
\label{DrivenSEPhi}
\end{equation}
and the boundary conditions read
\begin{equation}
\begin{array}{lll}
\Phi_{R}(k,0)&=&0,  \\
\Phi_{R}(k,r)&\sim  &({a^R})^{-1}{\cal A}_{R}\, u^{+}_{\ell}(\eta,kr).
\end{array}
\label{Phi_R-BC}
\end{equation}
When $kr\gg \ell(\ell+1)+\eta^2$ the outgoing Coulomb  wave $u^{+}_{\ell}$ has
the asymptotics~\cite{Abramowitz}
\begin{equation}
u^+_{\ell}(\eta,kr)\sim e^{\imi(kr-\ell\pi/2-\eta\log 2kr)},
\label{u^+as}
\end{equation}
and the asymptotics of $\Phi_{R}(k,r)$ becomes simpler
\begin{equation}
\Phi_{R}(k,r)\sim  ({a^R})^{-1}
{\cal A}_{R}\,e^{\imi(kr-\ell\pi/2-\eta\log 2kr)}.
\label{Phi_as}
\end{equation}
Furthermore, if $kR\gg \ell(\ell+1)+\eta^2$ and the asymptotics (\ref{aSRas})
applies, then the wave function asymptotics reduces to
\begin{equation}
\Phi_{R}(k,r)\sim
e^{-\imi\eta\log2kR}
{\cal A}_{R}\,e^{\imi(kr-\ell\pi/2-\eta\log 2kr)}.
\label{Phi_as-asR}
\end{equation}
The leading term of the amplitude ${\cal A}_R$ can be calculated from the value
of the wave function $\Phi_R(k,R)$ as
\begin{equation}
{\cal A}_{R}
\sim e^{2\imi \eta\log 2kR}\,\Phi_{R}(k,R)e^{-\imi(kR-\ell\pi/2)}.
\label{A_R-as}
\end{equation}
This formula gives the local representation for the amplitude.

The set of Eqs.~(\ref{DrivenSEPhi},\ref{Phi_R-BC},\ref{Phi_as-asR}) is our final
formulation of the Coulomb scattering problem based on the driven Schr\"odinger
equation with purely outgoing boundary conditions.

\section{Green's function formalism and integral representation for ${\cal A}_R$}
The differential equation (\ref{DrivenSEPhi}) with boundary conditions
(\ref{Phi_R-BC}) can easily be transformed into the integral Lippmann-Schwinger
equation
\begin{equation}
\Phi_R=-{\hat G}^RV_R{\hat{j_\ell}}-{\hat G}^RV_R\Phi_{R}.
\label{LSE}
\end{equation}
The kernel of the integral operator ${\hat G}^R$ is the Green's function
$$
G^R(r,r',k)=\langle r|(H_\ell+V^R-k^2-\imi 0)^{-1}|r'\rangle \,.
$$
It is defined by the standard expression 
\begin{equation}
G^{R}(r,r',k)=\frac{1}{k}\Psi^R(k,r_{<})U^R_{+}(k,r_{>}).
\label{GreenR}
\end{equation}
where $r_{>}(r_{<})=\max(\min)\{r,r'\} $, and $\Psi^R$ is the wave function
constructed in the preceding section. $U^R_{+}$ is the irregular solution to
Eq.~(\ref{SE-R}) with the asymptotics
as $r\to \infty$
\begin{equation}
U^R_{+}(k,r)\sim u^+_\ell (\eta,kr).
\label{U^R-as}
\end{equation}
This solution can be constructed by the matching procedure of the previous section.
On the interval $[0,R]$ the function $U^R_{+}(k,r)$ has the form of the superposition
of Riccati-Hankel~\cite{Abramowitz} functions,
\begin{equation}
U^R_{+}(k,r)=c^R\, {\hat h}^-_{\ell}(kr) + d^R\,  {\hat h}^+_{\ell}(kr).
\label{U^R-R}
\end{equation}
On the interval $[R,\infty)$ one sets $U^R_{+}(k,r)={\hat U}^R_{+}(k,r)$ (defined
in Eq.(\ref{UJost})). The continuity conditions of the function $U^R_{+}(k,r)$ and
its derivative over $r$ at the point $r=R$ yield for the coefficients $c^R$ and $d^R$
\begin{equation}
\begin{array}{l}
d^R=W_R({\hat U}^R_{+},{\hat h}^-_{\ell})/W_R({\hat h}^+_{\ell},{\hat h}^{-}_{\ell}), \\
c^R=W_{R}({\hat U}^R_{+},{\hat h}^{+}_{\ell})/W_R({\hat h}^{-}_{\ell},{\hat h}^{+}_{\ell}).
\end{array}
\label{cd}
\end{equation}
If the asymptotics (\ref{U^R-as}) can be applied  in Eqs. (\ref{cd}), and if
$kR\gg \ell(\ell+1)+\eta^2$, then the asymptotics for the coefficients $c^R$
and $d^R$ are 
\begin{equation}
c^R\sim 0, \ \ d^R\sim e^{-\imi \eta\log 2kR}.
\label{cd-as}
\end{equation}

Referring again to Eq.~(\ref{LSE}), the potential $V_R$ has finite range $R$, hence
for $r>R$ the integrals in Eq.~(\ref{LSE}) are taken on the finite interval $[0,R]$,
and in Eq.~(\ref{GreenR}) $r_>=r$ and $r_<=r'$ . Therefore, Eq.~(\ref{LSE}) for
$r>R$ is written as
\begin{equation}
\begin{array}{l}
\Phi_R(k,r)=-\frac{1}{k} a^R U^R_{+}(k,r) \\
\times \int^{R}_{0}dr'\, {\hat{j_\ell}}(kr')V(r')\left[ {\hat {j_{\ell}}}(kr')+\Phi_R(k,r')\right],
\end{array}
\label{chi2IE}
\end{equation}
where relation (\ref{Psi^R<R}) has been used. The asymptotics (\ref{U^R-as})
applied to Eq. (\ref{chi2IE}) leads to the desired integral representation for
the scattering amplitude
\begin{equation}
{\cal A}_R=-\frac{(a^R)^2}{k}\int^{R}_{0}dr'\, {\hat{j_\ell}}(kr')V(r')
\left[ {\hat {j_{\ell}}}(kr')+\Phi_R(k,r')\right].
\label{A_R-IR}
\end{equation}

\section{Application of exterior complex scaling to the driven Schr\"odinger equation\label{ECS}}
The success in solving the driven Schr\"odinger equation by the complex scaling
method depends on whether the driving term vanishes for complex values of the
coordinates. The driven Schr\"odinger equation formulation (\ref{DrivenSEPhi},\ref{Phi_R-BC})
perfectly meets this requirement since the potential in the right hand side is
of finite range. Another useful observation made from
representation (\ref{A_R-IR}) is that the scattering amplitude ${\cal A}_R$ is
completely determined by that part of solution $\Phi_R$, which is restricted on
the finite domain $0\leq r\leq R$. These features make the application of the ECS
method to the driven Schr\"odinger equation (\ref{DrivenSEPhi},\ref{Phi_R-BC})
ideally suited. Note that the ESC does not change the coordinates and the
solution in the interior domain transforming only the exterior part of the wave
function.

Let the ESC transformation operator
\begin{equation}
[W(Q,\alpha)\Psi](r)=\Psi(k,g_{Q,\alpha}(r)),\ \ Q\ge R
\label{W}
\end{equation}
be chosen in such a way that the transform $g_{Q,\alpha}(r)$ maps the interval
$[0,Q]$ into itself and the interval $[Q,\infty)$ into a path in the upper half
of the complex coordinate plane with the asymptotes $g_{Q,\alpha}(r)\sim e^{\imi\alpha}r$, $0<\alpha<\pi/2$. Then it follows from Eq.~(\ref{Phi_as}) that the W-transformed
solution to Eq.~(\ref{DrivenSEPhi}) $\Phi^W_R(k,r) = W(Q,\alpha)\Phi_R(k,r)$
has the asymptotics
\begin{equation}
\Phi^W_R(k,r)\sim (a^R)^{-1}{\cal A}_{R}
e^{\imi\left[k g_{Q,\alpha}(r)-\ell\pi/2-\eta\log 2k g_{Q,\alpha}(r)\right]}.
\label{chi^2g}
\end{equation}
This clearly shows that
\begin{equation}
\lim_{r\to \infty} \Phi^W_R(k,r) = 0.
\label{chi^2gzero}
\end{equation}
Let $H^W_{\ell}$ and $V^W$ denote the transformed operators
$W(Q,\alpha)H_{\ell}W^{-1}(Q,\alpha)$ and $W(Q,\alpha)VW^{-1}(Q,\alpha)$ respectively.
Then W-transformed driven Schr\"odinger equation (\ref{DrivenSEPhi}) takes the form
\begin{equation}
(H^W_{\ell} +V^W(r) - k^2)\Phi^W_R(k,r)=-V_R(r){\hat {j_\ell}}(kr). \label{WDSE}
\end{equation}
Furthermore, the W-transformed boundary conditions read
\begin{equation}
\Phi^W_{R}(k,0)=0, \ \ \Phi^W_{R}(k,r)\to 0,\ \ r\to \infty.
\label{WBC}
\end{equation}
The function $\Phi_R^W(k,r)$ coincides with $\Phi_R(k,r)$ for $r\leq Q$.
Therefore, if the equation (\ref{WDSE}) with zero boundary conditions (\ref{WBC})
has been solved, then the scattering amplitude ${\cal A}_R$ can be recovered
by the integral representation
\begin{equation}
{\cal A}_R=-\frac{(a^R)^2}{k} 
\int^{R}_{0}dr'\, {\hat{j_\ell}}(kr')V(r')
 \left[ {\hat{j_{\ell}}}(kr')+\Phi^W_{R}(k,r')\right], 
\label{WA_R-IR}
\end{equation}
or by the local representation
\begin{equation}
{\cal A}_{R}=
e^{2\imi \eta\log 2kR}\Phi^W_{R}(k,R)e^{-\imi(kR-\ell\pi/2)}.
\label{WA_R-as}
\end{equation}
The subsequent phase shift $\delta_R$ is calculated from the representation
(\ref{A_R}) and reconstructs the phase shift $\delta_{\ell}$ with the formula
\begin{equation}
\delta_{\ell}=\delta^R+\delta_{R}.
\label{delta_l}
\end{equation}


\begin{figure}[t]
   \centering
       \includegraphics[scale=0.75]{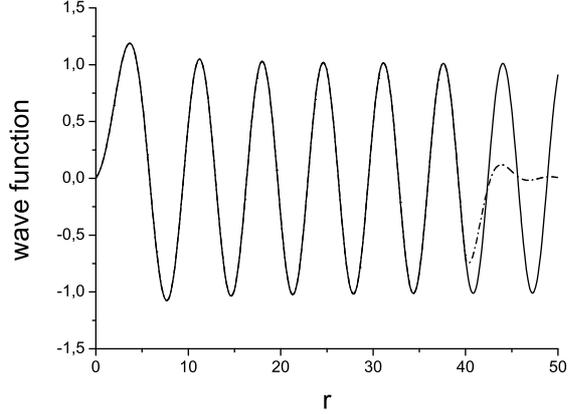}
   \caption{The Coulomb function $F_0(1/k,kr)$ (the solid line) and the real part of
   the function $(\Phi^W_R(k,r)+\hat{j}_0(kr))e^{\imi\delta^R}$ (the
   dashed-dotted line). The momentum $k=1$,
   the orbital momentum $\ell=0$, the radius $R=40$.}
   \label{fig1}
\end{figure}

\section{Numerical examples\label{Num}}
We have tested the results of the previous sections on some simple examples.
As a numerical method for the solution of Eqs.~(\ref{WDSE},\ref{WBC}), we have
chosen the FEM-DVR approach described in~\cite{R:number6}.
The parameters of the numerical approximation were chosen in such a way that the
numerical inaccuracies were negligible. For all the calculations presented, this
was already achieved with the equally spaced finite elements of the length 1 and
the Lobatto shape functions~\cite{R:number6} of the 10th degree.

In order to introduce the boundary conditions (\ref{WBC}) into the numerical
scheme, we chose the maximal radius $R_{max} > R$, where the second condition
(\ref{WBC}) is implemented. The results get quite stable with respect to the
maximal radius as soon as $R_{max}$ is considerably larger than $R$. Numerical
tests showed that our choice $R_{max}=1.25 R$ does not produce any
noticeably errors in the results.

While the radius $R$ and the exterior complex rotation radius $Q$ are allowed
to be different in our scheme, we have not yet found any advantages keeping them
distinct. Hence, in our calculations we choose $Q=R$. The calculations have also
showed that the specific choice of the ECS in Eq.~(\ref{W}) does not affect the
results. Therefore, we have here used the sharp ECS~\cite{r:SimonECS79}
\begin{equation} \label{ECS_sharp}
 g_{R,\alpha}(r) = \left\{
   \begin{array}{ll}
     r                       & \mbox{for } r \le R \\
     R+(r-R) e^{\imi \alpha} & \mbox{for } r > R
   \end{array}
 \right. .
\end{equation}
The rotation angle $\alpha$ was chosen to be 30 degrees.

Summarizing the description of our computational approach, we can say that the
only parameter affecting the results is the radius $R$. All other theoretical
and numerical parameters were chosen such that they do not influence the results.
We first solve Eqs.~(\ref{WDSE},\ref{WBC}) and compute the function $\Phi^W_R(k,r)$.
Then we find the amplitude ${\cal A}_{R}$ with the local representation
(\ref{WA_R-as}) or with the integral representation (\ref{WA_R-IR}).
Using Eq.~(\ref{A_R}) with $\delta^{R} = \eta\log 2kR-\sigma_{\ell}$, we calculate
the phase shift $\delta_R$ and finally reconstruct $\delta_\ell$ with the
relation (\ref{delta_l}).

We first compare our numerical results to the known analytic solution for the
pure Coulomb potential, $V(r)=2/r$. In Fig.~\ref{fig1}, we display the Coulomb
function $F_0(1/k,kr)$ together with the real part of the numerical solution
$(\Phi^W_R(k,r)+\hat{j}_0(kr))e^{\imi\delta^R}$ for the radius $R=40$ and the
momentum $k=1$. One can see that in the internal region $[0,R]$ the numerical
solution practically coincides with the exact one. Hence, it can be used to
determine the amplitude. Outside the internal region, for the complex rotated
coordinate, the function $\Phi^W_R(k,r)$ quickly approaches zero in accordance
with the second boundary condition in (\ref{WBC}). A detailed investigation
shows that the difference between the exact and numerical solutions in the
internal region is evenly distributed over the whole interval $[0,R]$ and
decreases when $R$ increases.

\begin{figure}[t]
   \centering
       \includegraphics[scale=0.75]{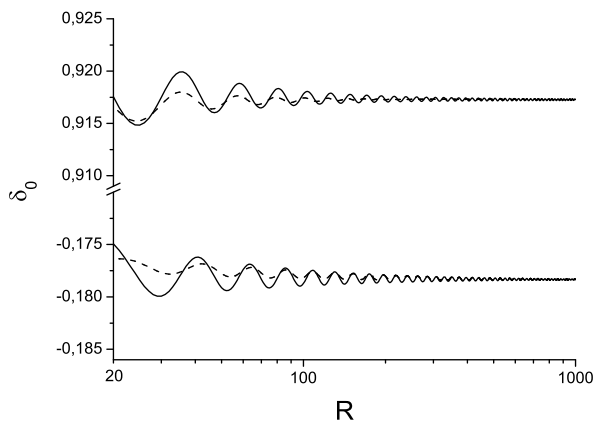}
   \caption{The phase shift $\delta_0$ as the function of the radius $R$. The
   integral representation (\ref{WA_R-IR}) (the solid line) and the local
   representation (\ref{WA_R-as}) (the dashed line) are shown. The upper lines
   correspond to the total potential (\ref{pot-example}), the lower lines
   correspond to the pure Coulomb potential. The momentum $k=3$.}
   \label{fig2}
\end{figure}

\begin{figure}[t]
   \centering
       \includegraphics[scale=0.75]{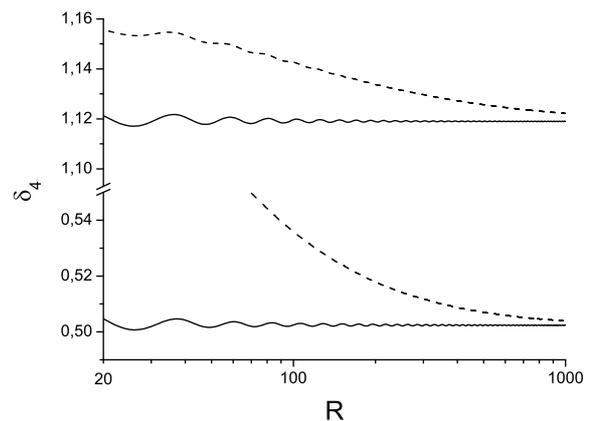}
   \caption{The phase shift $\delta_4$ as the function of the radius $R$. The
   notations are the same as in Fig.~\ref{fig2}.}
   \label{fig3}
\end{figure}

In Figs.~\ref{fig2} and \ref{fig3} we present the behavior of the local~(\ref{WA_R-as})
and the integral~(\ref{WA_R-IR}) representations of the amplitude with respect to
the radius $R$. We show results for both the pure Coulomb potential and the
potential with the short range term
\begin{equation} \label{pot-example}
 V(r) = 2/r + 15 r^2 e^{-r}.
\end{equation}
In Fig.~\ref{fig2} one can see that the accuracies for both Coulomb potential
$V(r)=2/r$ and potential given by Eq.~(\ref{pot-example}) are approximately the
same. For the chosen momentum, $k=3$, it is already as small as 1\% for the
relatively small radius $R=20$. The accuracy naturally depends on both the
radius $R$ and the momentum $k$ as we use the asymptotic expansion for
$kR \to \infty$. Thus, the bigger $kR$ is the higher is the accuracy.
The phase shift approaches the exact value when $R$ increases while the
convergence is not very fast. For zero angular momentum $\ell=0$,
the local and the integral representations give a comparable accuracy.

On comparing the data plotted in Figs.~\ref{fig2} and \ref{fig3}, there is an
important difference between results for zero angular momentum and a nonzero
angular momentum (here $\ell=4$). While the accuracy given by the integral
representation is about the same, the accuracy of the local representation gets
worse. This lack of accuracy is due to the condition $kR \gg \ell(\ell+1)+\eta^2$
which has to be satisfied in order to use equation~(\ref{Phi_as-asR}).
This means that we should consider larger values of $R$ to get the accurate result.
Another possibility could be to use the next order of the asymptotic expansion
with respect to $kR$ in Eq.~(\ref{Phi_R-BC}).

Finally, we have computed the s-wave scattering cross section for the potential
in (\ref{pot-example}) using the present code and a highly optimized logarithmic
derivative code based on the algorithm of Johnson~\cite{r:brjohnssson}.
The estimated relative accuracy of the logarithmic derivative results is smaller
than 0.001. The two cross sections agree within the same limits.

\section{Conclusions\label{Concl}}
In this note we have presented a mathematically sound formulation which describes
how the exterior complex scaling can be applied to the system with the long-range
(Coulomb) interaction to compute scattering quantities. This formulation does not
require any knowledge of the exact Coulomb solutions and can, therefore, be
generalized to the three-body scattering problem with the charged particles. In the
absence of the Coulomb potential, $\eta=0$, our formulation also presents the
rigorous justification for the ECS method~\cite{number1}. This is achieved by
reducing the solution of the scattering problem with a long-range (including Coulomb)
potential to the solution of a boundary value problem (\ref{DrivenSEPhi},\ref{Phi_R-BC}).
This reduction is exact for an arbitrary value of the radius $R$, and is also
supplied with the exact integral representation (\ref{A_R-IR}) for the scattering
amplitude. For large values of $R$, the asymptotics can be
used in order to explicitly calculate the boundary conditions (\ref{Phi_R-BC}).
Furthermore, the boundary problem 
is combined with the ECS technique as the potential in the right hand
side of Eq.~(\ref{DrivenSEPhi}) is the finite-range potential. The numerical
implementation of the theory shows both good accuracy and high efficiency of
the approach.

\acknowledgments
M.V.V. and E.Y. are very thankful to Swedish Institute for support.
The work of M.V.V. and N.E. is supported by a grant from the Swedish Research Council.
The work of E.Y. and S.L.Y. was partially supported by the Russian Foundation for
Basic Research grant 08-02-01115. S.L.Y. is thankful to Stockholm University for
the support of his visit made possible under the bilateral agreement on
cooperation between Stockholm University and St.Petersburg University.

\end{document}